\documentclass[10pt]{aastex}

\title{\large On the Need for Deep Mixing in AGB Stars of Low Mass}

\author{M. Busso\altaffilmark{1}, S. Palmerini \altaffilmark{1}, E. Maiorca\altaffilmark{1}, S. Cristallo\altaffilmark{2,3}, O. Straniero\altaffilmark{3,4}, C. Abia\altaffilmark{2}, R. Gallino\altaffilmark{3,5}, \& M. La Cognata\altaffilmark{6,7}}

\altaffiltext{1}{Dipartimento di Fisica, Universit\`{a} di
Perugia, and INFN, Sezione di Perugia, Italy; maurizio.busso@fisica.unipg.it}
\altaffiltext{2}{Departamento de F{\'\i}sica Te\'orica y del Cosmos, Universidad de Granada, Spain}
\altaffiltext{3}{INAF, Osservatorio di Teramo, Italy }
\altaffiltext{4}{INFN, Sezione di Napoli}
\altaffiltext{5}{Dipartimento di Fisica Generale, Universit\`{a} di Torino}
\altaffiltext{6}{Dipartimento di Metodologie Fisiche e Chimiche per l'Ingegneria, Universit\`{a} di Catania}
\altaffiltext{7}{INFN, Laboratori Nazionali del Sud, Catania, Italy}

\begin{document}
\begin{abstract}
The photospheres of low-mass red giants show  CNO isotopic abundances that are not satisfactorily accounted for by canonical stellar models. The same is true for the measurements of these isotopes and of the $^{26}$Al/$^{27}$Al ratio in presolar grains of circumstellar origin. Non-convective mixing, occurring during both Red Giant Branch (RGB) and Asymptotic Giant Branch (AGB) stages is the explanation commonly invoked to account for the above evidence. Recently, the need
for such mixing phenomena on the AGB was questioned, and chemical anomalies usually attributed to them were suggested to be formed in earlier phases. We have
therefore re-calculated extra-mixing effects in low mass stars for both the RGB and AGB stages, in order to verify the above claims. Our results contradict them; we actually confirm that slow transport below the convective envelope occurs also on the AGB. This is required primarily by the oxygen isotopic mix and the $^{26}$Al content of presolar oxide grains. Other pieces of evidence exist, in particular from the isotopic ratios of carbon stars of type N, or C(N), in the Galaxy and in the LMC, as well as of SiC grains of AGB origin. We further show that, when extra-mixing occurs in the RGB phases of population I stars above about 1.2 $M_{\odot}$, this consumes $^3$He in the envelope, probably preventing the occurrence of thermohaline diffusion on the AGB. Therefore, we argue that other extra-mixing mechanisms should be active in those final evolutionary phases.
\end{abstract}

\section{Introduction}

Evolved low-mass stars show photospheric CNO isotopic ratios that in many cases
are not reproduced by stellar evolutionary codes. In the past
years it was recognized by many authors \citep{boot,wbs,char,sb99}
that these chemical anomalies derive from transport
mechanisms linking the envelope to zones where partial
H-burning occurs. We call these phenomena ``extra-mixing'' or ``deep mixing'' throughout this report. \citet[][hereafter NBW03]{nol} presented a parametric study of such mixing episodes, suitable to account for the CNO abundances measured in presolar grains of AGB origin. The adopted formalism was based on two parameters, namely the rate of mass transport ($\dot{M}$)
and the temperature ($T_P$) of the deepest zones reached by
the circulation. It was also demonstrated that important composition changes can occur without introducing feedbacks on the stellar luminosity, provided $T_P$ is kept low enough (typically, $\Delta \log~ T = \log~T_{\rm H} - \log~T_P \gtrsim$ 0.08$-$0.1, where $T_{\rm H}$ is the temperature at which the maximum energy of the H-burning shell is released).

Subsequently, physical models for extra-mixing have been explored, which avoid the difficulties previously found with rotationally-induced mechanisms \citep{pal}. In particular, hydrodynamical models of diffusive processes induced by variations of the mean molecular weight $\mu$ \citep{ss69}, called {\it thermohaline diffusion} \citep{cz07}, were presented by \citet{egg1,egg2}. They showed that $^3$He burning into $^4$He and two protons successfully induces the required
$\mu$ inversion ($\Delta \mu/\mu \simeq -$ 10$^{-4}$), thus driving mixing episodes. Complementarily, \citet{bwnc}, \citet{nord} and \citet{den} suggested that extra-mixing might be driven by magnetic buoyancy, in a dynamo process operating below the envelope.

The model by NBW03 referred explicitly to AGB stars, which are known to be the parents of most presolar grains \citep{zin1}. Extra-mixing in these objects was considered as necessary by various authors \citep{hoppe,zinner,heck,nit2}. However, a recent paper \citep[][hereafter KCS10]{kar} sheds doubts on this requirement, suggesting that slow mass circulation on the RGB might be sufficient. In order to solve the above dilemma, we have to look at the observational data, asking whether there is any clear requirement for extra-mixing on the AGB. Our answer will be that this requirement exists and comes primarily from oxide presolar grains \citep{nit1,choi,clay,nit2}. Then one must look for a similarly clear constraint from stars, and we find it in the carbon isotope ratios of C(N) giants. Other pieces of evidence (from Ba stars, C-enhanced metal poor stars, etc.), although relevant, will not be addressed here for reasons of space. In order to demonstrate all this we generalize the calculations by NBW03, extending them to cover also RGB stages and including two more values of the initial mass. In Section 2 we compare our results for RGB phases with constraints coming from presolar oxide grains of group 2. In Section 3 we integrate this by considering extra-mixing also in AGB stages; critical tests come again from presolar oxide grains (of group 2, but also of group 1). The evidence from O-rich AGB stars is also discussed. Finally, in Section 4 we comment on the information coming from C(N) stars and we derive preliminary conclusions, also addressing the physical mechanisms necessary to drive the transport.

\section{Extra-mixing in RGB Phases}

For the sake of comparison with previous works we perform parametric calculations, using the free parameters $\dot M$ and $T_P$, as in NBW03. We adopt
the same NACRE compilation for reaction rates and we compute our extra-mixing results as post-process calculations, starting from detailed stellar models by \citet{stran}. Masses in the range 1.2 $\le$ $M/M_{\odot}$ $\le$ 2 for $Z = Z_{\odot}/2$ are considered (only the case of a 1.5 $M_{\odot}$ star was analyzed in NBW03). During RGB stages, we introduce extra-mixing after the ``bump'' of the luminosity function (BLF), when the H-burning shell erases the chemical discontinuity left behind by the first dredge up (FDU); hence envelope abundances do not change from FDU to this moment \citep[see, e.g.,][]{char96}. Details on the general RGB physics can be found in \citet{cb} and in \citet{pal}. Our procedure was illustrated by \citet{palm1} and is based on the integration of a set of partial differential equations describing the abundance changes due to nucleosynthesis and to the upward and downward transport across the integration grid. The rate of mass circulation $\dot M$, defined in NBW03, is assumed as a free parameter. In agreement with the above paper the fractional areas occupied by the upward and downward streams are assumed to be the same.

\begin{figure*}[t!!]
\centering {\includegraphics[height=6.8cm]{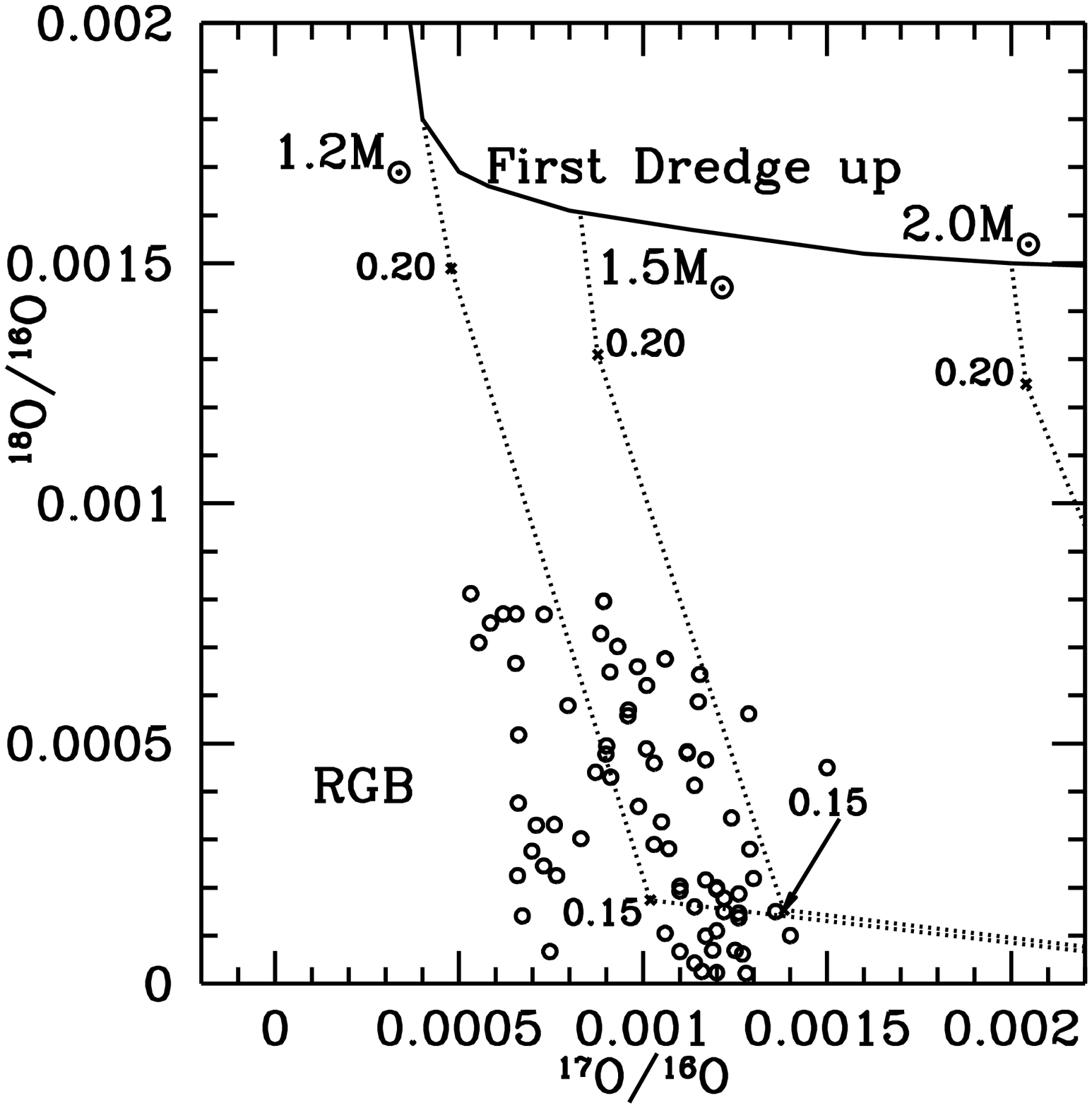}
\includegraphics[height=6.8cm]{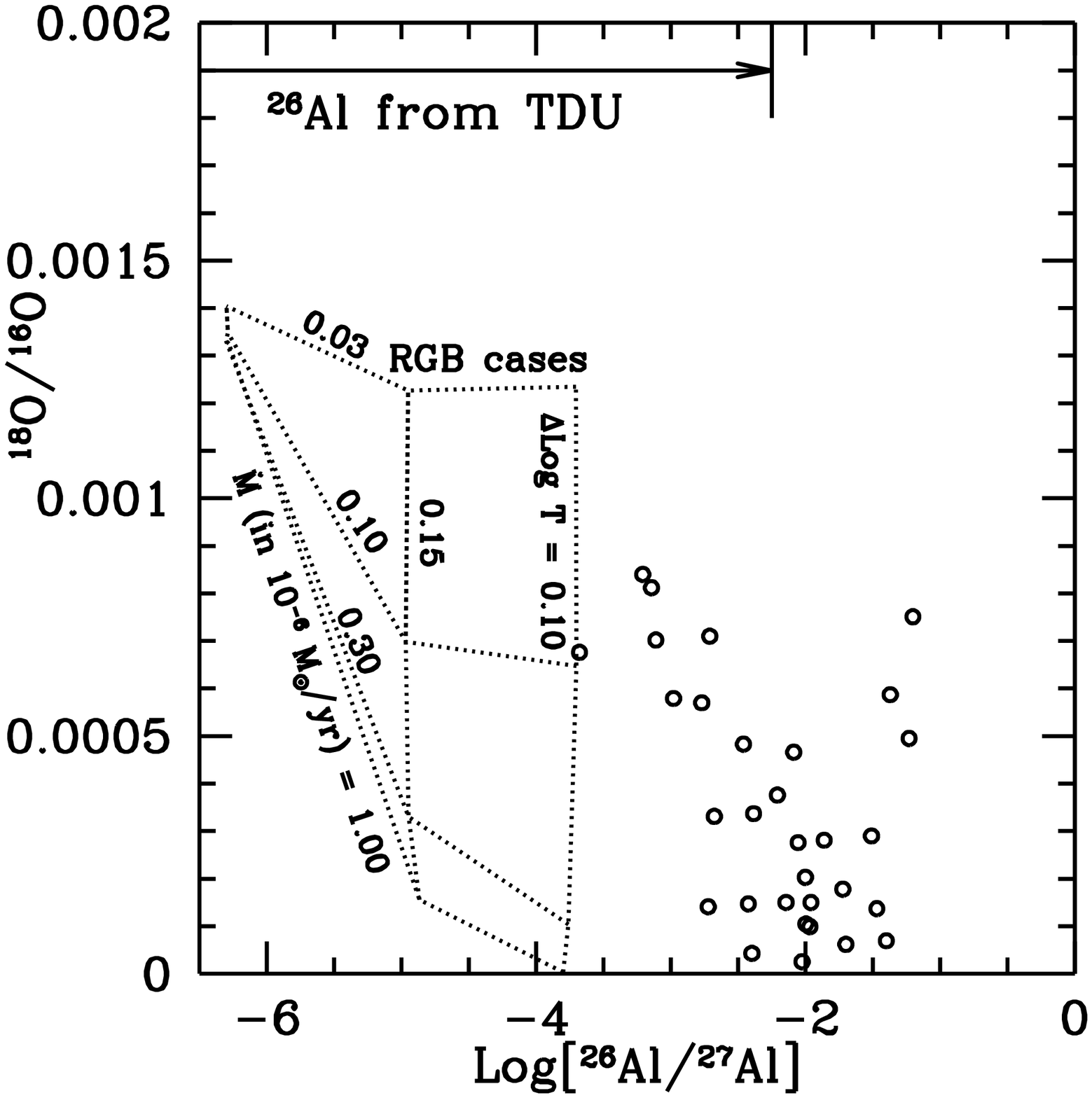}}
\caption{Left: oxygen isotopes in presolar oxide grains of group 2 (open circles), as compared to deep mixing models in RGB phases. Trajectories with $\Delta \log T < 0.15$ lie on an asymptotic line of low slope, ending at $^{17}$O/$^{16}$O $\simeq$ 0.003, typical of CNO equilibrium at relatively low temperatures. Right: the ratio $^{18}$O/$^{16}$O in oxide grains of group 2 as a function of their inferred $^{26}$Al/$^{27}$Al ratio. The grid of model lines shows the region covered by extra-mixing on the RGB, for various $\dot M$ and $T_P$ values, as indicated by the labels. The observed $^{26}$Al data are not explained by RGB phases.  The arrow shows the amount of $^{26}$Al subsequently provided by the third dredge up. See text for details.} \label{one}
\end{figure*}

In the cases discussed here, the results are essentially controlled by the path integral of reaction rates over the trajectory of the transport. They are shown,
for the RGB phases of our three model stars, in Figure \ref{one}. The left panel shows our predictions for the oxygen isotopic ratios, as compared to measured values in presolar oxide grains of group 2 \citep{nit1, nit2}. These are the grains whose composition is attributed to extra-mixing phenomena, because of the extensive depletion of $^{18}$O. The continuous track in the plot identifies the composition at FDU for various stellar masses. The dotted lines display the isotopic ratios obtained with efficient mixing ($\dot M$ = 10$^{-6}$ $M_{\odot}/$yr) continued from the BLF to the RGB tip. Along each line, the displacement from the FDU composition increases for decreasing values of  $\Delta \log~ T$; the points corresponding to some values of this parameter are shown with labels on the model lines.  The results for $\Delta \log~T \simeq$ 0.10 and for very low
$^{18}$O/$^{16}$O lay on an asymptotic trajectory with low slope pointing to a high $^{17}$O/$^{16}$O ratio (0.003). This is determined by the equilibrium abundances of oxygen isotopes in CNO cycling: at the low RGB temperatures ($T \le 3.5\cdot10^7$ K) the concentration of $^{17}$O is very high (see NBW03, Figure 8). It is clear that extra-mixing on the RGB alone does not account for the measurements in the most $^{18}$O-poor grains. Indeed, it would require a much higher content of  $^{17}$O than observed. The right panel of Figure \ref{one} adds more evidence by including the Al isotopic ratio. The grid of dotted curves refers to extra-mixing calculations on the RGB for a 2 $M_{\odot}$ model: the results are however typical of the whole range of low mass stars. The curves roughly going from left to right are for $\dot M$ values of 0.03, 0.1, 0.3, 1 in units of 10$^{-6}$ $M_{\odot}/$yr. The higher is the $\dot M$ value, the larger is the $^{18}$O depletion, as indicated by the labels. The almost vertical lines, instead, refer to different values of $\Delta \log~T$ (0.2, 0.15, 0.1, as indicated). It is clear that extra-mixing on the RGB might cover only Al isotopic ratios below 10$^{-4}$; this is an irrelevant contribution, especially when considering that much more $^{26}$Al (up to $^{26}$Al/$^{27}$Al $\simeq$ 5$-$6$\times$10$^{-3}$) is subsequently dredged up after thermal pulses on the AGB (even in the absence of further extra-mixing episodes on the AGB itself). These simple comparisons make clear that deep mixing on the RGB alone is inadequate to explain the constraints coming from presolar grains of circumstellar origin.

\section{Extra-mixing in AGB Phases}

In  \citet{ir83} and \citet{bgw} one can find generalities for AGB stars. Below their envelopes, conditions of chemical homogeneity are established when the downward envelope expansion called {\it third dredge-up} (TDU), which follows the {\it thermal pulses} of the He shell, reaches down to the H-He discontinuity. Most of the interpulse periods of AGB phases are therefore suitable for the occurrence of extra-mixing, which can therefore operate for a total duration of about 10$^6$ yr. The spread shown by presolar grains and by AGB stars in the C, O and Al isotopic ratios suggest that the details of the transport vary from star to star (NBW03).
\begin{figure*}[t!!]
\centering {\includegraphics[height=6.8cm]{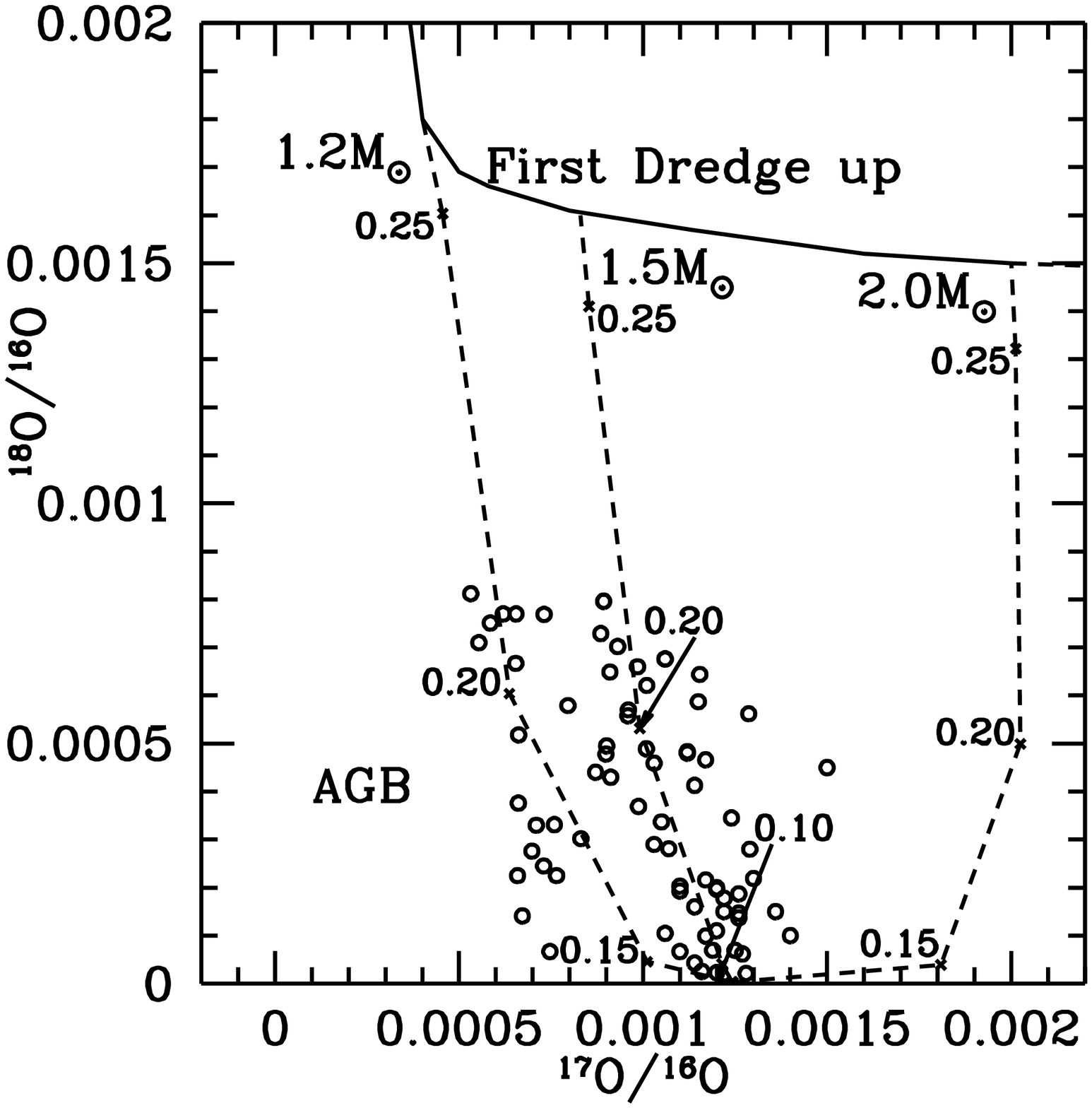}
\includegraphics[height=6.8cm]{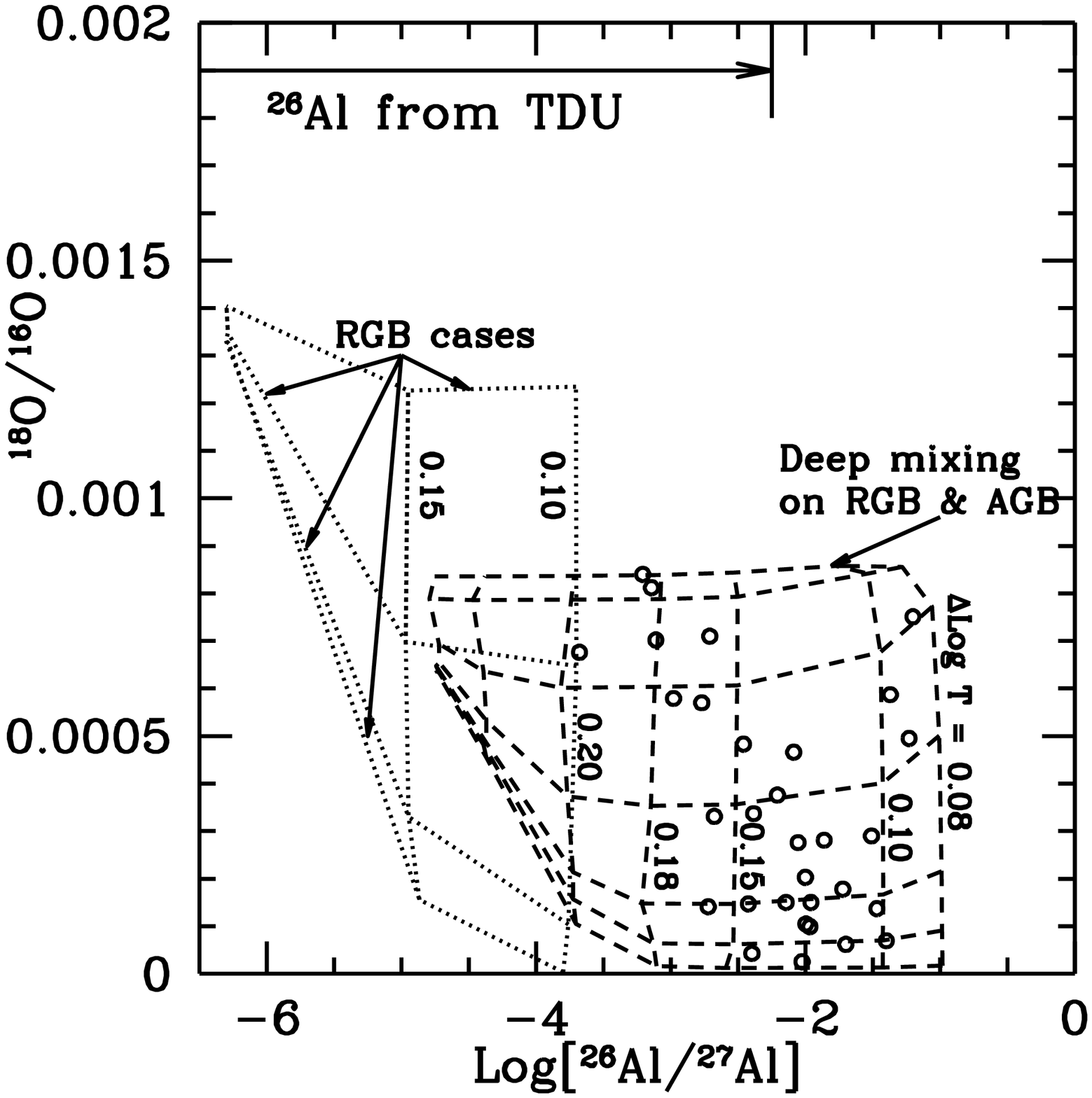}}
\caption{Left: oxygen isotopes in presolar oxide grains of group 2 (open circles), as compared to models of extra-mixing in AGB phases, for stars of different masses. Any previous phenomenon of circulation on the RGB would be hidden in the plot, as its tracks would be indistinguishable from the  upper parts of the AGB model lines. Right: the ratio $^{18}$O/$^{16}$O in oxide grains of group 2 (open circles) as a function of their inferred $^{26}$Al/$^{27}$Al ratio. The grid of dashed lines refer to cases where extra-mixing occurs on both the RGB and the AGB; the arrow shows the contribution from TDU (see text).} \label{two}
\end{figure*}

Figure \ref{two} (left panel) shows the oxygen isotopic ratios reachable by extra-mixing in the interpulse phases of AGB stars. Dashed lines show the effects of efficient mixing ($\dot M = 5 \cdot 10^{-6}$ $M_{\odot}$/yr); again the displacement from the FDU composition grows for increasing depth (hence temperature) reached by the transport ($\Delta \log~T$ values are indicated on the curves). Model predictions converge to $^{17}$O/$^{16}$O $\simeq$ 0.0012, characteristic of CNO equilibrium at the moderately high temperatures found above the H-burning shell in low-mass AGB stars ($T \simeq 5-6\cdot 10^7$ K). The grain isotopic ratios cluster around this region and are well reproduced. The right panel of Figure \ref{two} adds the evidence provided by Al isotopes. Dashed lines show the region covered by models of deep mixing, if it occurs in both the RGB and the AGB phases of a 2 $M_{\odot}$ star with $Z=Z_{\odot}/2$. The previous activation of the same phenomenon on the RGB is considered by assuming initial isotopic abundances in the range obtained from the computations of Section 2. Again, the grid of model lines shows cases with various $\dot M$ values (0.01, 0.03, 0.1, 0.3, 1, 3 and 5, in units of 10$^{-6}$ $M_{\odot}$/yr, higher $\dot M$ values corresponding to higher $^{18}$O consumption). The almost vertical lines of the grid refer to $\Delta~\log~T$ = 0.25, 0.22, 0.20, 0.18, 0.15, 0.10, 0.08 (see the labels in the plot). The RGB sequences of Figure 1 (right panel) are shown as dotted lines for comparison. The data of group 2 grains are well explained by the models, suggesting that they were formed in stars that experienced extra-mixing in both RGB and AGB phases.

It is well known that another mixing phenomenon occurs on the AGB, for intermediate mass stars (IMS: $M \ge 4.5 - 5 M_{\odot}$). It is driven by the convective envelope extension down to the H-burning layers, in a process called Hot Bottom Burning (HBB). The isotopic mix of presolar oxide grains cannot however be explained by this mechanism; this was made clear in the works by \citet{sb99}, \citet{lug} and \citet{ili}. In particular, \citet{ili} revised the $^{16}$O(p,$\gamma$) reaction rate, essentially confirming the NACRE value but reducing the uncertainty. On this basis, they found an equilibrium $^{17}$O/$^{16}$O ratio in their IMS model of 2.52$^{+0.88}_ {-0.76}\times10^{-3}$, incompatible with presolar grain data. They concluded that ``there is not clear evidence to date for any stellar grain origin from {\it massive} AGB stars'' \citep[see also][Figure 7]{sb99}. The measurements of Figure \ref{two} cluster around an $^{17}$O/$^{16}$O ratio of 0.0012, as typical of the shell H-burning temperatures in AGB stars of masses up to about 2 $M_{\odot}$. Both lower and higher temperatures (as found in RGB and HBB conditions, respectively) would bring the model tracks outside the area of the measured data. This fact, when coupled with the abundances of $^{26}$Al in the grains, offers compelling evidence in favor of deep mixing in the AGB phases of low mass stars. Our results would not change significantly by adopting the reaction rate for $^{17}$O+p from \citet{chafa}, as done by KCS10: the differences in the equilibrium $^{17}$O/$^{16}$O values between this choice and the NACRE one amount (at maximum) to 20\%.

One may further notice that, among the eight optically-visible, moderately-evolved MS and S stars observed by \citet{har}, four (50\%) have $^{18}$O/$^{16}$O ratios below 5$\times$10$^{-4}$, and two (25\%) around 2$\times$10$^{-4}$. Despite the low statistics and large error bars of stellar observations, this compares well with the family of oxide grains requiring extra-mixing. As an example, in the Saint Louis database (http://presolar.wustl.edu/$\sim$pgd/) one finds 76 grains of group 1 and 79 grains of group 2 with  $^{18}$O/$^{16}$O ratios lower than permitted by FDU. For 35\% of the whole sample this ratio is below 5$\times$10$^{-4}$ and for 17\% it is below 2$\times$10$^{-4}$. Among group 2 grains alone these numbers become 68\% and 33\%, respectively. A straightforward interpretation is that visible O-rich AGB stars represent an intermediate population, experiencing extra-mixing more effectively than the {\it average} of the grains, but less effectively than the {\it extreme} (group 2) grains. This is encouraging. For a direct comparison, one would need oxygen isotopic ratios in evolved, extinct AGB stars, which are the parents of most grains. They are however difficult to observe. Conversely, stellar observations show an anti-correlation of $^{18}$O/$^{16}$O with C/O ratios that is reasonable to expect from AGB evolution, but that cannot be verified in oxide grains.

Extra-mixing seems therefore to be a common property of low mass AGB stars; and it must be a normal occurrence also on the RGB, as revealed by the low carbon isotope ratios of first-ascent red giants \citep{lr81, cos,shetr}. Notice, instead, that only few evolved stars are Li-rich \citep{abia2}. However, producing Li requires mixing at high rates, due to the short $^7$Be lifetime; such fast mixing episodes might be found only rarely.

\section{Discussion and Conclusions}

 Our results indicate that extended mixing on the AGB is required, when considering oxygen isotopes and $^{26}$Al/$^{27}$Al ratios in presolar grains. \cite[In particular, the oxygen isotopes maintain their crucial role in constraining stellar physics, as early noticed by][]{dearb}. Stellar evidence also points to the same conclusion. This is so in particular for carbon isotopic ratios in C(N) stars. Indeed, while extra-mixing on the RGB is sufficient to explain such ratios (averaging around 60) in the C(N) star sample by \citet{lamb}, this is no longer true for the stars of \citet{ab02}. These authors agree rather well (on average) with \citet{lamb} for the sources in common; however, they also identified a number of new C(N) stars (25\%) with $^{12}$C/$^{13}$C ratios $\simeq 10 - 40$. Mainstream SiC grains provide the same evidence (for 23\% of them, in the St Louis database, the carbon isotope ratio is below 40, while the average value is around 60, as for C(N) stars). \citet{ab02} found that even assuming, at the beginning of the AGB phase, $^{12}$C/$^{13}$C $\simeq$ 12, as due to extra-mixing on the RGB, final C-isotope ratios in excess of 43 were always obtained at C/O=1, unless new extra-mixing episodes on the AGB were included. Hence, C(N) giants and SiC grains confirm that extra-mixing occurs on the AGB; this is required also by their oxygen isotopes \citep{har1,kaha}. The requirements from O-rich stars and grains suggest however a more extended processing than for C(N) stars. As these last have, on average, a larger mass than MS-S giants \citep{gb08}, the extra-mixing efficiency in population I seems to decrease with increasing stellar mass. For lower metallicities, extra-mixing effects are known to be enhanced \citep{ch98,grat00}.

In the KCS10 approach, C(N) stars and SiC grains with $^{12}$C/$^{13}$C ratios below about 33 (see also their Table 1) cannot be explained, so that the constraints of SiC grains and carbon stars with low carbon isotope ratios are not reproduced. Note that the difference between the values found by \citet{kar} and by \citet{ab02} for the  $^{12}$C/$^{13}$C ratios in C-rich stars (without extra-mixing) is due to different choices for this same ratio at the end of the RGB phase, and for C/O. KCS10 also claim that they ``cover most of the range of observational data points''. They refer to C stars in the Galaxy and in the LMC clusters NGC1846 and NGC1978. Although their Figure 2 might give this impression, this is not the case for the clusters. In fact, O-rich and C-rich stars in each cluster represent an evolutionary sequence, whose C/O and $^{12}$C/$^{13}$C ratios must be explained together. In contrast with this requirement, KCS10 need, for explaining C(N) stars, initial C/O and $^{12}$C/$^{13}$C values of 10 and 0.32; for explaining O-rich AGB stars they instead need 22 and 0.23, respectively (see their Figure 2). KCS10 actually admit their inability at explaining the two constraints together; this failure shows that their approach is not correct, at least for NGC 1846. Conversely, a scenario accounting for both O-rich and C-rich stars was presented by \citet{leb08}, introducing a moderate extra-mixing on the AGB. These authors also suggested an increased efficiency of extra-mixing for increasing values of the envelope C/O ratio, because the bottom of the convective envelope becomes progressively closer to the H-burning shell while the star climbs the AGB. For the same reason, the extra-mixing efficiency is expected to increase for AGB stars of low metallicity, as the convective envelope becomes hotter for them \citep{cris}.

The situation of NGC1978 is definitely more puzzling \citep{lederer}: in that case, a fit to C-rich stars requires the concomitant absence of extra-mixing processes during both the RGB and AGB phases. Thus, it is hard to find a theoretical recipe suitable to reproduce the isotopic ratios of AGB stars in NCG1978 (both O-rich and C-rich) without invoking an ad-hoc solution for this peculiar cluster \citep[see discussion in][]{lederer}.

\begin{figure}[t!!]
\centering{\includegraphics[height=6.0cm]{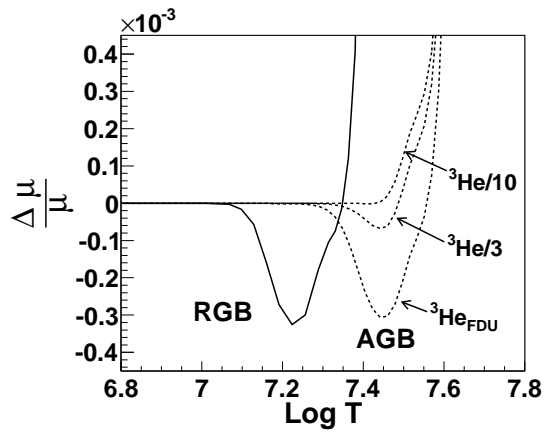}}
\caption{Relative variation of the molecular weight ($\Delta \mu /\mu$) for a 1.5 M$_{\odot}$ star with a metallicity $Z = Z_{\odot}/2$, in the layers where $^3$He burns. The inversion, present on the RGB, is reduced or erased on the AGB when $^3$He has been previously consumed (see labels for consumption factors).} \label{three}
\end{figure}

Finally, we would like to comment on the physical origin of extra-mixing on the AGB. A popular mechanism is today thermohaline diffusion \citep{egg1,egg2}. However, it is unlikely that it occurs on the AGB of population I stars, when any deep mixing has been previously active on the RGB. Indeed, the envelope abundance of $^3$He, whose burning drives the mixing through a $\mu$ inversion, would be considerably reduced.  This is shown in Figure \ref{three}, where we plot the mean molecular weight across the radiative region in RGB and AGB phases, for our 1.5 M$_{\odot}$ star. While a $\mu$ inversion of $\Delta \mu /\mu \simeq -3 \times 10^{-4}$ (continuous line) is in fact driven by $^3$He burning on the RGB, the AGB cases (dashed lines) are more critical. They refer either to no destruction of $^3$He on the RGB ($^3$He$_{\rm FDU}$), or to a consumption by factors 3 or 10 (this last case corresponds to the findings by KCS10). It is evident that, for the masses and metallicities considered here, the $\mu$ inversion is either strongly reduced or erased if $^3$He has been previously consumed. As extra-mixing (hence $^3$He consumption) on the RGB of galactic disk stars is required by observations \citep{cos,shetr}, the conditions for thermohaline diffusion might be suppressed in their AGB stages \citep[see also][]{can}. Other mechanisms should therefore be looked for, including magnetic buoyancy \citep{bwnc}. Note that thermohaline mixing was found to occur on both the RGB and the AGB (first interpulses) in a low-metallicity 1-$M_{\odot}$ star \citep{stan}. This is due to the known higher inventory of $^{3}$He in very low-mass stars \citep{dearb1}.

{\bf Acknowledgements}. We acknowledge useful comments from G.J. Wasserburg and from two very constructive referee reports. We are indebted to E. Zinner and coworkers for maintaining the on-line repository of presolar grain abundances from which we took the measured data (http://presolar.wustl.edu/~pgd/). C. Abia acknowledges partial support by the Spanish grant AYA2008-04211-C02-02

\end{document}